# An ILP formulation to optimize flood evacuation paths by minimizing pedestrian speed, length and effort

*Fabrizio Marinelli[1]\*, Andrea Pizzuti[1], Guido Romano[2], Gabriele Bernardini[2]\*\*, Enrico Quagliarini[2]*

[1] DII Department, Università Politecnica delle Marche – Ancona, Italy
[2] DICEA Department, Università Politecnica delle Marche – Ancona, Italy
\* fabrizio.marinelli@staff.univpm.it
\*\* g.bernardini@staff.univpm.it

**Abstract**. This document presents an Integer Linear Programming (ILP) approach to optimize pedestrian evacuation in flood-prone historic urban areas. The model aims to minimize total evacuation cost by integrating pedestrian speed, route length, and effort, while also selecting the optimal number and position of shelters. A modified minimum cost flow formulation is used to capture complex hydrodynamic and behavioral conditions within a directed street network. The evacuation problem is modeled through an extended graph representing the urban street network, where nodes and links simulate paths and shelters, including incomplete evacuations (deadly nodes), enabling accurate representation of real-world constraints and network dynamics.

Integer Linear Programming (ILP) techniques are increasingly used in emergency evacuation planning to identify optimal pedestrian routing and shelter allocation under complex spatiotemporal, behavioral and infrastructural constraints [1,2]. Problems are typically modeled as minimum cost flows over directed graphs, where binary decision variables encode pedestrian movements along network edges [3]. Constraints can ensure flow conservation, limit the number of selected nodes, and consider accessibility conditions. By encoding factors such as travel time, length, and capacity limitations into a unified optimization framework, ILP supports the development of efficient and scalable evacuation strategies, making it particularly suitable for urban scenarios exposed to flood risk [4,5].

Specifically, ILP consist of three components: 1) a set of integer decision variables codifying the quantitative features of decision strategies; 2) a linear objective function indicating the optimization criteria for evaluating solution quality; 3) a set of linear constraints defining the feasible region, i.e., the variables' values that describe allowed decision strategy.

Based on the classical linear formulation of the minimum cost flow problem, i.e., the problem of sending the flow supplied from a given set of source nodes of a network $G$ to a given set of sink nodes of $G$ at the minimum cost, an Integer Linear Program (ILP) for solving the optimal evacuation problem is presented in this section. Under the classical assumptions of time-constant and conservative flows -the former states that the network is in a stationary state and the latter that the total amount of flow entering each node $u$, including the quantity $p_u > 0$ supplied from outside, must be equal to the total amount of flow leaving the node including the quantity $p_u < 0$ absorbed from the outside- we study the evacuation problem modeling the routes of pedestrians as a flow throughout a given network. In our case, the network $G = (V, E)$, where $V$ is the set of nodes and $E$ is the set of links, is uncapacitated and oriented, i.e. each link $(u,v)$ has a capacity $u_{uv}$ unbounded (which is consistent with the quantity of simulated pedestrians) and can be crossed only by a flow from $u$ to $v$ and not vice-versa.

The optimal evacuation problem can be stated as: given a street network, locate at most $M$ shelters and find the pedestrian evacuation routes of minimum total cost. Graph $G$ is the representation of the street network of the idealized urban area, where the set $V$ of nodes are the starting points of pedestrians (every node on the map) and $V_s \subseteq V$ are the subset of potential shelters (nodes circled in red).

Nodes in $V$ are associated with reference points of the streets and of the square. Links in $E$ (in green) model both the segments of streets and elementary routes in the square. Due to the considered application, directed links model the possible directions of the pedestrian evacuation, which generally are from the source of risk and cannot be upstream [6,7]. It follows that the streets parallel to the river, which therefore can be traveled in both directions, are represented by bidirectional links (or equivalently by non-oriented links). On the other hand, perpendicular and oblique streets induce oriented links in the graph moving away from the river, with the only exception for the "dead-end" streets furthest from the river that can only be left by moving upstream to remain within the boundaries. The cost $c_{uv}$ of each link $(u,v)$ is computed according to the pedestrian speed, stability, and effort on the flooded streets by considering the hydrodynamic conditions of the floodwater spreading within the urban area.

According to the case study description, the instance of the graph considered in this paper presents $|V|$ = 41 staring points, $|E|$ = 130 links, and $|P|$ = 240 pedestrians that, before the event, have been randomly distributed among the starting points. As a result, 86 pedestrians have been allocated to the square, 11 pedestrians to each street perpendicular to the river, and 4 pedestrians to each street parallel to the river [7].

Moreover, $|V_s|$ = 12, i.e., 12 nodes of $G$ are potential shelters, basically corresponding to (quasi-)dry areas (i.e., $DV \approx 0$ m$^3$/s), chosen according to the hydrodynamic conditions reported in a recent previous study [7]. In particular, potential shelters located in the midpoints of the parallel streets simulate the entrance to indoor safe areas, while that in the square simulate an outdoor safe area.

Dummy nodes and connections are added to $V$ and $E$ in order to formulate the optimal evacuation problem on $G$ in terms of a particular minimum cost network flow problem on a new graph $G'$. In Figure A 1, given the graph $G$ in (a) which is a subgraph of the street network associated with the urban area, (b) depicts the graph $G'$ obtained by adding the blue nodes and the dotted links. In details:

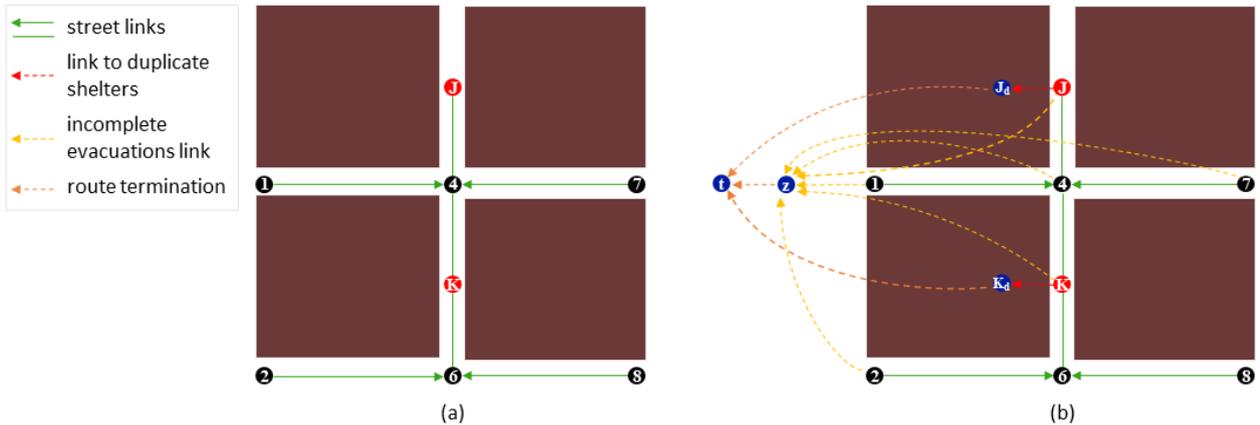

**Figure A 1**: (a) subgraph of the street network; (b) subgraph obtained considering the dummy nodes (in blue) and related link (dotted). Legend on the left (orientated links are the green arrows, non-orientated links are the green lines).

- For each potential shelter $u \in V_s$, a new node $w$ with $p_w = 0$ is introduced, along with an oriented link $(u, w)$ with $c_{uw} = 0$ (dotted red links in (b)). The set $V_d$, indeed a duplicate of $V_s$, allows the incoming flow at node $u \in V_s$ to exit the network from the same node $u$ at zero cost, i.e., it allows the pedestrians whose starting point is also a potential shelter to select that shelter without moving.
- A new node $z$ with $p_z = 0$ is added and then linked at cost $c_{uz} = C$ (dotted ochre links in (b)) to each starting node $u \in V \setminus V_d$. The node $z$ is needed to balance the flow of pedestrians who cannot complete the evacuation. The constant $C$ is sufficiently large to ensure that, in any optimal solution, a pedestrian chooses the destination $z$ if and only if it cannot reach any potential shelter.

- A sink node $t$ with $p_t = -|P|$ is added and connected by oriented links $(u,t)$ of cost $c_{ut} = 0$ to all the nodes in $V_d \cup \{z\}$ (dotted orange links in (b)). The sink $t$ is used to balance the whole incoming/outgoing flow in the network since it absorbs, at no cost, all the $|P|$ units of flow coming from either the nodes in $V_d$ (successful evacuations) or from node $z$ (uncompleted evacuations).

zIn summary, the extended graph $G'$ of the urban area has $|V| = 55$, $|V_d|=12$, and $|E|= 130$ and, given its size, we do not explicitly report it. The notation used for the ILP is shown in Table A 1.

| PARAMETERS | MEANING |
| --- | --- |
| $V = \{1, ..., |V|\}$ | set of nodes, including the dummy nodes $t, z$, and the dummy nodes in $V_d$ |
| $V_s \subseteq V$ | subset of potential shelter nodes |
| $E = \{1, ..., |E|\}$ | set of links |
| $P = \{1, ..., |P|\}$ | set of pedestrians |
| $p_u^k$ | = 1 if the pedestrian $k$ starts the evacuation from the node $u$ and = 0 otherwise |
| $p_u = \sum_{k \in P} p_u^k$ | total number of pedestrians starting evacuation from node $u \neq t$ |
| $p_t = -|P|$ | total flow absorbed by sink node $t$ |
| $c_{uv}$ | cost of the link $(u,v)$ between nodes $u$ and $v$ |
| $C$ | upper bound to the costs $c_{uz}$ with $u \in V \setminus V_d$ |
| $N$ | maximum number of evacuation points |
| VARIABLES | MEANING |
| $f_{uv}^k \begin{cases} = 1 \\ = 0 \end{cases}$ | if the pedestrian $k$ travels the link $(u,v)$ <br> otherwise |
| $y_u \begin{cases} = 1 \\ = 0 \end{cases}$ | if the potential shelter $u$ is chosen <br> otherwise |

Table A 1: ILP model – notation

The parameter $p_u^k$ is set to one if the node $u$ is the starting point of the pedestrian $k$, and 0 otherwise. Thus, $p_u = \sum_{k \in I} p_u^k$ is the total amount of pedestrians starting the evacuation from node $u$. Moreover, we recall that $M$ is the maximum number of nodes of $V_s$ that can be selected as shelters.

The model is defined on the vectors of decision variables $\mathbf{f} \in \{0,1\}^{|P|\times|E|}$ and $\mathbf{y} \in \{0,1\}^{|V_d|}$: $f_{uv}^k = 1$ if and only if the pedestrian $k$ walks the link $(u,v)$, whereas $y_u = 1$ if and only if the node $u$ is selected as shelter (actually, $y_u$ refers to the node in $V_d$ which is the duplicated of the node $u$ in $V_s$).

About the ILP constraints, the classical flow-balancing equalities of the minimum cost flow problem read as:

$$\sum_{\substack{v \in V: \\ (v,u) \in E}} f_{vu}^k + p_u^k = \sum_{\substack{v \in V: \\ (u,v) \in E}} f_{uv}^k \quad \forall k \in P, u \in V \quad (2)$$

Enforcing the network to be conservative, the constraint ensures that a pedestrian $k$ enters a node $u \in V$, because starting from it ($p_u^k = 1$) or arriving from an incoming link $(v, u)$, must leave it through one of the outgoing links $(u, v)$. Since the sink node $t$ has no outgoing links, we can sum up the pedestrians and obtain:

$$\sum_{k \in P} \sum_{u \in V_d \cup \{z\}} f_{ut}^k = -p_t \quad (3)$$

which imposes, by recalling that $p_t = -|P|$, all the pedestrians to reach the sink.

Further inequalities are required to control the shelters' selecting variables **y**:

$$\sum_{k \in P} \sum_{\substack{v \in V: \\ (v,u) \in E}} f_{vu}^k \leq |P| \cdot y_u \quad \forall u \in V_d \quad (4)$$

If any pedestrian $k$ reaches the shelter located at node $u$ (actually at its duplicated node in $V_d$), then the shelter $u$ must be selected ($y_u = 1$). In that case, i.e., when $y_u = 1$, the constraint does not restrict the capacity of the shelter, being $|P|$ an upper bound to the number of pedestrians that can reach the node $u$. If otherwise $y_u = 0$, then the evacuation from the shelter $u$ is not allowed.

The last constraint just bounds from above the number of selected shelters to $M$:

$$\sum_{u \in V_d} y_u \leq M \quad (5)$$

A minimum total cost flow on $G'$ describes an optimal evacuation plan. Then, the objective function is given by:

$$\min \sum_{k \in P} \sum_{(u,v) \in E} c_{uv} f_{uv}^k \quad (6)$$

The above ILP models the case where a single decision-maker (the coordinator of the evacuation operations) computes evacuation routes in a centralized fashion and assigns them to pedestrians. Moreover, minimizing the sum of costs, the objective function (6) assumes a collaborative dynamic of groups where a single pedestrian could accept a difficult evacuation route if this choice helps many other pedestrians. However, alternative objective functions, e.g., minimizing the cost of the most difficult pedestrian's path, could be implemented in order to equally rescue all the pedestrians.

The above single objective function actually implements a hierarchical optimization between two objectives: minimizing the number $Z$ of the pedestrians unable to complete the evacuation and, for the same $Z$, minimizing the total cost of the successful evacuations. Such a hierarchy is obtained by choosing the cost $C$ of the links $(u, z)$ large enough. Other than prioritizing the intrinsic value of life, the above hierarchical optimization avoids typical drawbacks of the multi-objective optimization [8] such as the extensive computation of the optimal Pareto frontier, or the sensitivity analysis on weights associated to the convex combination of terms.

If $G$ is a strongly connected and uncapacitated network, the problem boils down to the optimal selection of shelters, which in any case is not trivial since $|P| \gg M$ in general. Indeed, given a selection $\bar{y}$ of shelters, each pedestrian has a feasible evacuation route ($G$ strongly connected), and such route corresponds to the minimum cost route (e.g., the shortest path) from the pedestrian starting point to $t$ ($G$ uncapacitated). Therefore, for a given selection $\bar{y}$ of shelters, the optimal solution is simply the collection of the shortest routes of the pedestrians. Nevertheless, this work considers that $G$ is only weakly connected. Thus, not all the selections $\bar{y}$ of shelters, or even none of them, guarantee a successful evacuation route for each pedestrian. Indeed, optimal solutions of the case study have $Z > 0$.

Clearly, the shortest evacuation routes are no longer independent of each other in the case of capacitated networks due to the presence of saturated links that may force one or more pedestrians to deviate from their own optimal route. This naturally increases the total cost of solutions, both by augmenting the cost of completed evacuations and the occurrences of uncompleted evacuations.

The proposed ILP is solved through the commercial package IBM-CPLEX[1]. The experiments aimed to identify optimal solutions according to three different link costs $c_{uv}$, evaluated in combination with a range of values for the shelter availability $M$.

For $i$ {s, q, c}, Table A2 reports the details specifying the behavioral approach for the route choice leading to the settings of the link costs: the first approach chooses the *shortest* routes to minimize the evacuation length, the second one looks at the *quickest* routes to reduce the evacuation time, the third one minimizes the evacuation effort through *cheapest* routes.

| *i* | *Route choice strategy* | *Minimized parameter* | *Link cost c* |
|---|---|---|---|
| s | Shortest | Evacuation length | S [m] |
| q | Quickest | Evacuation time | T [s] |
| c | Cheapest | Evacuation effort | DV·S [m3/s] |

Table A 2: Evacuation strategies for route choice and their computation method depending on the link cost.

As a result, 36 scenarios ($i$, $M$) are defined by combining the link cost type $I$ {s, q, c} and the number of available shelters $M$ {1, 2, …,12}. An optimal solution $[\mathbf{f}^*, \mathbf{y}^*]_{(i,M)}$ has been computed for each scenario ($i$, $M$), where $Z_{(i, M)}$ is the minimum number of pedestrians unable to complete the evacuation, and $C_{(i, M)}$ is the total cost given by Eq. (6). The optimal solution describes the selection $\mathbf{y}^*_{(i,M)}$ of shelters, the evacuation route $\mathbf{f}^*_{(i,M)}$ from the starting node to the shelters of each pedestrian, the number $Z_{(I,M)}$ of pedestrians unable to complete the evacuation, and an array $\boldsymbol{C}_{\text{AVG}(i,M)} = [S_{\text{AVG}}; T_{\text{AVG}}; DVS_{\text{AVG}}]_{(i,M)}$ listing the mean cost values restricted to the $|P| - Z_{(i,M)}$ successful evacuation routes.

---

[1] https://www.ibm.com/products/ilog-cplex-optimization-studio/cplex-optimizer (last accessed: 18/03/2025)